\documentclass[11pt]{amsart}                                                  
\usepackage{hyperref}
\usepackage{url}
\usepackage[svgnames]{xcolor}
\newtheorem*{anti*}{Antithesis}

\newtheorem{terms}{Terminology}
\newtheorem{thesis}{Thesis} 

\newcommand{\Ar}{\medskip\noindent\textbf{A:\ }}

\renewcommand\phi{\varphi}

\newcommand{\Qn}{\medskip\noindent\textbf{Q:\ }}

\title[Church-Turing thesis]{Unconstrained Church-Turing thesis
cannot possibly be true}
\author[Yuri Gurevich]{Yuri Gurevich\\ 
{\scriptsize University of Michigan}}
\address{Computer Science and Engineering\\
University of Michigan\\
Ann Arbor, MI  48109, U.S.A}
\email{gurevich@umich.edu}
\thanks{This is an expanded and dialogized version of an October 21, 2018 talk at Microsoft Research Redmond. The characters A and Q are the author and his former student Quisani.}

\begin{document}
\begin{abstract}
The Church-Turing thesis asserts that if a partial strings-to-strings function is effectively computable then it is  computable by a Turing machine. 

In the 1930s, when Church and Turing worked on their versions of the thesis, there was a robust notion of algorithm. These traditional algorithms are known also as classical or sequential. In the original thesis, effectively computable meant computable by an effective classical algorithm. Based on an earlier axiomatization of classical algorithms, the original thesis was proven in 2008.

Since the 1930s, the notion of algorithm has changed dramatically. New species of algorithms have been and are being introduced. We argue that the generalization of the original thesis, where effectively computable means computable by an effective algorithm of any species, cannot possibly be true.

\end{abstract}
\maketitle

\section{Introduction}
\label{sec:intro}

\Qn
There is a discussion about the Church-Turing thesis at the Computer Science Theory StackExchange \cite{StackExchange}. It involves your paper \cite{G188} with Nachum Dershowitz where you prove the thesis. Peter Shor is skeptical about it: 
\begin{quote}
``The Dershowitz-Gurevich paper says nothing about probabilistic or quantum computation. It does write down a set of axioms about computation, and prove the Church-Turing thesis assuming those axioms. However, we're left with justifying these axioms. Neither probabilistic nor quantum computation is covered by these axioms (they admit this for probabilistic computation, and do not mention quantum computation at all), so it's quite clear to me these axioms are actually false in the real world, even though the Church-Turing thesis is probably true.''
\end{quote}
What do you say?

\Ar 
The Church-Turing thesis asserts that if a string function is effectively computable then it is  computable by a Turing machine. 
Here a string function is a partial function from strings in a finite alphabet to strings in a finite alphabet.

In the 1930s, when Church and Turing worked on their versions of the thesis, there was a robust notion of algorithm. 
These traditional algorithms are known also as classical or sequential algorithms. It is this notion of algorithm which is axiomatized in \cite{G141}. In the original thesis, the effective computability of a string function means that it is computable by an effective classical algorithm. It is that original thesis which is proven in the Dershowitz-Gurevich paper \cite{G188}. 

Since the 1930s, many new species of algorithms have been introduced, and the notion of algorithm continues to evolve \cite{G209}. 
Apparently Peter Shor thinks that we pretend to prove the unconstrained version of the thesis, for the algorithms of all species, and that the unconstrained thesis is true.

\Qn But surely the validity of the thesis is not restricted to the classical algorithms.

\Ar I believe that the thesis can be proven for a number of well-understood species of algorithms, in particular for algorithms in the quantum circuit model. But the unconstrained version of the thesis cannot possibly be true.

\Qn Please explain.

\section{The original Church-Turing Thesis}
\label{sec:orig}\mbox{}\\[-6ex]

\Ar Let me quickly revisit the original thesis; details and relevant references are found at \cite{G188}. I will address the unconstrained version later.

\subsection{Classical algorithms}
\label{sub:seq}\mbox{}
 
The 1930s notion of algorithm was robust. People recognized algorithms when they saw them. 
These algorithms compute in steps, one step after another, and the steps are of bounded complexity \cite{Kolmogorov}.
Various names are used today for those algorithms: traditional, classical, sequential. 

\Qn None of the three names seems perfect to me.
Tradition changes with time.  ``Classical'' may mean merely not quantum. ``Sequential'' seems consistent with unbounded complexity of steps.

\Ar This is true. To distinguish between bounded and unbounded complexity of steps, we spoke about small-step and wide-step algorithms in \cite{G164}. But even that neglects the distinction between classical algorithms and algorithms  interacting with their environment as well as the distinction  between classical and learning algorithms.  

\begin{terms}
\emph{Classical algorithms} are algorithms in the sense of the 1930s (rather than merely non-quantum algorithms).
\end{terms}

Classical algorithms were analyzed and axiomatized in \cite{G141}. The analysis and axiomatization were refined in \cite{G201}, mostly because the original analysis abstracted from details of intra-step computation.

\subsection{The thesis}
\label{sub:thesis}\mbox{}

There are many equivalent formulations of the Church-Turing thesis. The Dershowitz-Gurevich article \cite{G188} was published in a logic journal. There, respecting logic tradition, we formulated the thesis in terms of partial recursive functions. Here, respecting computer science tradition, we  formulate the thesis in terms of Turing machines. 

\begin{terms}\label{pro:terms}\mbox{}\rm
\begin{itemize}
\item A \emph{string function} is a partial function from strings in a finite alphabet to strings in a finite alphabet.
\item A string function is \emph{Turing computable} if it is computable by some Turing machine.
\end{itemize}
\end{terms}

Now we can formulate the Church-Turing thesis succinctly. Here is a generic version of the thesis which leaves open what is meant by effective computability.

\begin{thesis}[Generic Church-Turing thesis]\label{the:gen}
If a string function is effectively computable  then it is Turing computable.
\end{thesis}

\noindent
The approriate version of the original/classical thesis is this:

\begin{thesis}[Classical Church-Turing thesis]\label{the:cls}
If a string function is computed by an effective classical algorithm then it is Turing computable.
\end{thesis}

\Qn What does it mean exactly that an algorithm computes a string function?

\Ar Without loss of generality, we can define this in a way convenient for our purposes. An algorithm $A$ computes a string function $f$ if 
\begin{itemize}
\item[-] inputs of $A$ are strings $x$ in the input alphabet of $f$,
\item[-] if $f$ is defined at $x$ then the computation of $A$ on $x$ eventually converges and outputs $f(x)$, 
\item[-] if $f$ is not defined at $x$ then the computation of $A$ on $x$ produces an error message or diverges, i.e., goes on forever.
\end{itemize}
Notice that this definition abstracts from limited resources. In the real world, a computation of an algorithm $A$ on input $x$ may break because we ran out of time or money or something else. 

\Qn What does it mean that an algorithm $A$ is effective? 

\Ar An algorithm $A$ is effective if, given sufficient resources, the computation of $A$ on any input $x$  can be carried out in the real world.

\Qn Show me some noneffective algorithms.

\subsection{Noneffective classical algorithms}
\label{sub:ineff}\mbox{}

\Ar One example is Euclid's algorithm for lengths.
You know Euclid's algorithm for natural numbers; given two natural numbers, the algorithm computes their greatest common divisor. Euclid used a similar algorithm for lengths; today we can think of lengths  as nonnegative real numbers. Given two lengths, the algorithm finds their greatest common divisor if the two lengths are commensurable and diverges otherwise. 

Another example is Gauss Elimination algorithm for real numbers.

\Qn In both cases, reals can be approximated by rationals as closely as desired, and the computation on rationals can be carried out effectively.

\Ar This is true though the approximating algorithm will be much more involved, and there are some subtleties. For example, two reals  may or may not be commensurable while any two rationals are commensurable.  Besides, noneffective classical algorithms may be more abstract. For example, Gauss Elimination works over every field.

\Qn Neither of the two noneffective algorithms computes a string function.

\Ar Oracle algorithms, which compute string functions, may be and often are noneffective. In particular, a Turing machine with an appropriate oracle solves the halting problem for oracle-free Turing machines. 

\Qn Using oracles looks like cheating. 

\Ar But it may be useful. Turing used oracles machines already in 1939 \cite{Turing39}. 

\subsection{Proving the original thesis}
\label{sub:proof}\mbox{}

\Qn Your proof of the thesis appeared only on 2008 \cite{G188}. How come that the thesis wasn't proven earlier?

\Ar One reason for that could be that it is easier to axiomatize all classical algorithms rather than only effective ones. The proof of the thesis builds on the axiomatization of classical algorithms in \cite{G141}

\Qn But people could think of all classical algorithms earlier on. 

\Ar It was natural to restrict attention to effective algorithms. Turing for example ignores noneffective algorithms completely in his thesis paper \cite{Turing}. 

With time, software grew more involved, and software specifications started to use oracles and even work with genuine reals.

\Qn Did you axiomatize algorithms with an eye on proving the Church-Turing thesis?

\Ar No, not at all. I introduced abstract state machines (originally called evolving algebras) and posited a thesis that every algorithm is an abstract state machine \cite{G103}. The purpose of the axiomatization in \cite{G141} was to prove the new thesis for classical algorithms.

Later, Nachum Dershowitz and I extended that axiomatization  with an axiom saying essentially that there is no funny stuff in the initial state of the algorithm. This allowed us to derive the Church-Turing thesis \cite{G188}.

\section{Unconstrained Church-Turing thesis}
\label{sec:gen}

\Ar Let's formulate the unconstrained thesis more explicitly.

\medskip
\begin{thesis}[Unconstrained Church-Turing thesis]\label{the:un}
If a  string function is computed by any effective algorithm whatsoever then it is Turing computable.
\end{thesis}

\noindent
Now I am ready to posit my antithesis.

\medskip
\begin{anti*}
The unconstrained thesis cannot possibly be true.
\end{anti*}

\Qn How do you justify the Antithesis?

\Ar Let me give you three arguments.

\subsection{A moving target}
\label{sub:evol}\mbox{}

\smallskip
My first  argument is related to the evolution of the notion of algorithm. The notion of algorithm keeps evolving and getting more liberal \cite{G209}. This makes it a moving target. 

In that sense it is analogous to the notion of number. We have already many species of numbers, e.g.,
\begin{itemize}
\item integers, rationals and reals,
\item complex numbers and algebraic numbers,
\item quaternions, octonions, sedenions, 
\item ordinal numbers and cardinal numbers,
\item non-standard numbers, introduced by Abraham Robinson, and surreal numbers, introduced by John Conway.
\end{itemize}
And surely new species of numbers will be introduced. One should be careful about claiming that a  property is common to all species of numbers.

\newpage
Similarly, we have already many species of algorithms, e.g.,
\begin{itemize}
\item sequential and parallel algorithms, 
\item nondeterministic algorithms,
\item real-time and analog algorithms, 
\item randomized and probabilistic algorithms, 
\item distributed algorithms, 
\item quantum algorithms, 
\item biology-inspired algorithms,
\item learning algorithms.
\end{itemize}
And surely new species of algorithms will be introduced. 
One should be careful about claiming that a  property is common to all species of algorithms.  

\Qn I cannot think of any intrinsic property of all numbers.
Some but not all numbers are quantities, some but not all numbers represent orderings. Yet, as far as I know, addition and multiplication are defined for all species of numbers. This property seems to survive the introduction of new species of numbers; it may be common to all species of numbers, present and future. 
By analogy, there should be properties common to all species of algorithms, present and future. It is possible a priori that the validity of the Church-Turing thesis is such a property. 

\Ar I will argue that this is not the case. 

\subsection{Engineering}
\label{sub:eng}\mbox{}

Classical algorithms are mathematical objects. Large real-world algorithms of today are engineering systems. Typically they perform tasks and provide services, but sometimes they compute string  functions as well.
My second argument in favor of the Antithesis is that, in the case of large real-world algorithms, the Church-Turing thesis is sort of trivially true and therefore uninteresting.

Consider for example a popular industrial compiler for some common programming language, e.g.\ C++, which  has been written by many people. Typically such a compiler runs on numerous platforms, but for simplicity let's fix a platform. The compiler computes a string function: 
In comes a source code, and out goes an object code or an error message. 

\Qn But are compilers algorithms? 

\Ar Semantically, any software product is an algorithm, in my opinion.  
But notice that the generic Thesis~\ref{the:gen} does not use the term algorithm.  It is about effective computability. We could reformulate the unconstrained Thesis~\ref{the:un} by replacing ``algorithm'' with a term that sounds more inclusive, e.g.\ ``computing system.'' 

\Qn A question arises whether the Church-Turing thesis holds for real-world algorithms --- or computing systems --- like compilers.

\Ar Any real-world compiler accepts only finitely many source programs. It doesn't accept source programs which are too long or too involved. 
The function computed by the compiler is finite and therefore recursive.

\Qn This is disappointing. The thesis is true but uninteresting.  Can we  abstract from limited resources in this case?

\Ar Any popular industrial compiler is updated from time to time. Some bugs are fixed, and the new version may accept some source programs which had not been accepted earlier. Assume that the compiler will be updated forever and that  there are infinitely many source programs $P$ such that some version of the compiler accepts $P$.  

\Qn For our purposes, there is an ambiguity problem with such a continuously developing compiler. It does not compute a single-valued string function. Different versions may treat the same source program differently.

\Ar  Furnish every application of (any version of) the compiler with a unique identity. Formally, the identity is a part of compiler input, and this way we solve the ambiguity problem. But the compilation process does not use the identity.

\Qn The resulting  string function does not seem to be Turing computable which challenges the Church-Turing thesis. But people may disagree that a continuously developing compiler is an algorithm or even a computing system. 

\Ar This brings me to my third argument.

\subsection{Changing attitude}
\label{sub:attitude}\mbox{}

Let me start with another example and then formulate my third argument in favor of the Antithesis.
 
Consider Google Translate \cite{GoogleTranslate} and fix some source language, say English, and some target language, say Russian.  An English text (a query) is translated into Russian. I presume that every application of Google Translate is furnished with a unique identity. Such an application can be seen as a pair $(X,Y)$ where $X$ is a so-called unique query, i.e.\ an English text with the unique identity, and $Y$ is the resulting translation to Russian.  All such pairs $(X,Y)$ form a function which I will call GT. The abstraction of unlimited resources renders GT infinite. 

\Qn I do not like when you apply the abstraction of unlimited resources to real-world systems. Companies come and go, and so do their tools.
But at least in this case the abstraction looks more natural than in the compiler case. Even though Google Translate is continuously learning and thus continuously changing, or maybe because of this, it is more naturally perceived as one entity than a sequence of compiler versions.

\Ar Do you think that GT is Turing computable?

\Qn Surely not. Let's suppose that a Turing machine $T$ computes GT. Then  $T$ ``knows'' how  English and Russian will develop, in particular what English slang will emerge and how it will be translated to Russian with its own new slang. This is absurd.

\Ar Would you consider GT effectively computable?

\Qn Hmm, GT  is certainly computable in practice. As a frequent user of Google Translate, I know that it works. Furthermore, it works fast, almost in no time. 
The translation may be poor but this is beside the point. 

If effective algorithms are algorithms that work in practice then Google Translate is an effective algorithm. I am somewhat bothered that Google Translate is so different from algorithms of my college days. It is being trained on huge data. Its program keeps changing.  What do you think?

\Ar My opinion is that practically computable functions like GT are effectively computable.  My third argument in favor of the Antithesis is that this opinion will become more and more common. 
There is an informative analogy between the following two questions.
\begin{itemize}
\item Are practically computable functions effectively computable?
\item Can machines think?
\end{itemize}
Here is an instructive quote of Turing \cite[\S6]{Turing50}:
\begin{quote}
``The original question, `Can machines think?' I believe to be too
meaningless to deserve discussion. Nevertheless I believe that at
the end of the century the use of words and general educated opinion
will have altered so much that one will be able to speak of machines
thinking without expecting to be contradicted.'' 
\end{quote}

\noindent
 Life is a better opinion-changer than arguments.


\subsection{Finale}
\label{sec:final}\mbox{}

\Qn Let me review your arguments to get a better overall picture. Your first argument in favor of the Antithesis is that the notion of algorithm is a moving target and therefore one should be cautious with universal claims about all algorithms. 

Your second argument is that, in the case of large real-world programs, the Church-Turing thesis is at best uninteresting. This erodes the thesis somewhat but does not demolish it. 

It is the third argument that is most damaging to the thesis. String functions like the translation function GT are practically computable but not Turing computable. So why you don't claim that the unconstrained thesis is plainly false?


\Ar My opponents may argue that Google Translate is not really an algorithm because, in addition to a given text in the source language, huge data has been used to train Google Translate, 
or because Google Translate keeps changing its program. 
In my view, the continuing progress will render these counterarguments less and less convincing.

\Qn Do you expect that it will be recognized eventually that the unconstrained Church-Turing thesis is false?

\Ar This  outcome is possible. Notice, however, that the thesis requires the unlimited-resources abstraction. In the case of Google Translate, this abstraction requires that Google Translate works forever.  In the real world, the unlimited-resources abstraction is absurd and, I expect, will be viewed as such. The unconstrained thesis itself will be considered meaningless. 

In either case, whether the unconstrained thesis is considered false or meaningless, it is not true, and so the Antithesis holds.

\subsection*{Acknowledgments}\mbox{}

Many thanks to Andreas Blass for most useful discussions throughout my work on this dialog. 

I am grateful also to my colleagues who took time to comment on the penultimate version of the dialog: Cris Calude, Anuj Dawar, Pierre Lescanne, Leonid Levin, Naphtali Rishe, Alexander Shen, Volodya Vovk.

\end{document}